\documentclass[runningheads]{llncs}

\usepackage{graphicx}

\usepackage{color}
\usepackage{ulem}

\definecolor{mygreen}{rgb}{0.42,0.56,0.14}

\begin{document}
\title{{\sl Maximus}: a Hybrid Particle-in-Cell \\ Code 
for Microscopic Modeling \\ of Collisionless Plasmas}
\titlerunning{{\sl Maximus}}
\author{J.A. Kropotina\inst{1,2} \and
A.M. Bykov\inst{1,2} \and \\
A.M. Krassilchtchikov\inst{1} \and
K.P. Levenfish\inst{1}
}
\authorrunning{J.A. Kropotina et al.}
\institute{Ioffe Institute, St.~Petersburg, Russia \and
Peter the Great St. Petersburg Polytechnic University, St.~Petersburg, Russia \\
\email{email: juliett.k@gmail.com}\\
}
\maketitle              
\begin{abstract}
A second-order accurate divergence-conserving hybrid particle-in-cell 
code {\sl Maximus} has been developed for microscopic modeling of
collisionless plasmas. The main specifics of the code include 
a constrained transport algorithm for exact conservation of magnetic field divergence, 
a Boris-type particle pusher, a weighted particle momentum deposit 
on the cells of the 3d spatial grid, an ability to model multispecies plasmas, 
and an adaptive time step.
The code is efficiently parallelized for running on supercomputers by means 
of the message passing interface (MPI) technology; an analysis of parallelization 
efficiency and overall resource intensity is presented.
A {\sl Maximus} simulation of the shocked flow in the Solar wind is shown to agree
well with the observations of the Ion Release Module (IRM) aboard the Active 
Magnetospheric Particle Tracer Explorers interplanetary mission.

\end{abstract}
\section{Introduction}

Collisionless plasmas are ubiquitous in space on a vast range
of scales from the Earth magnetosphere to galaxy clusters -- the
largest bound objects in the universe. The growing amount and enhancing
quality of multiwavelength observational data require adequate models
of the observed objects and structures to be developed and confronted
with the data.
The required modeling is a very complicated task, as it has to account 
with sufficient accuracy and detalization for physical processes
occurring on a very broad range of spatial and temporal scales, such as
self-consistent acceleration of charged particles along with the
evolution of the underlying bulk plasma flows and multiscale 
electromagnetic fields.

As substantial energy of the bulk flows can be converted into
accelerated particles, whose dynamical role and back-reaction on the
structure of the flows is significant, the modeling can not be performed
within the frame of magnetic hydrodynamics -- it has to be done
on a microscopic kinetic level, and even further, within 
a particle-in-cell approach, where dynamics of individual particles
(ions and electrons) is modelled on a spatial grid with piecewise-constant 
values of the electromagnetic field.

Hybrid models are a special class of particle-in-cell models of plasma 
flows, where the ions are treated as individual particles, while the
electrons are considered as a collisionless massless fluid 
\cite{Matthews94,Lipatov,WinskeO93}. Such an approach allows one 
to resolve nonlinear collisionless effects on ion scales, 
at the same time saving substantial computational resources 
due to gyro-averaging of fast small-scale motions of the electrons. 
Hybrid codes are employed to study various astrophysical
processes, such as evolution of ion-ion instabilities 
or shocked collisionless plasma flows  
\cite{quest88,giacalone97,burgess12,caprioli_2014a,caprioli_2014b,caprioli_2014c,MarcowithEtal2016},
with the advantage of much greater time and spatial scales compared to those
of the full particle-in-cell approach.

Here we present a detailed description of the numerical scheme 
and parallelization technique of {\sl Maximus\footnote{
The code is named after out late colleague, Maxim Gustov (1978-2014),
who developed its first version in 2005.}}, a second-order
accurate divergence-conserving hybrid particle-in-cell code intended
for modeling of nonlinear evolution of collisionless plasma flows
with shocks. Previous versions of the code are briefly described
in \cite{kropotina_2015,kropotina_2016}. The main features of 
{\sl Maximus} are: a divergence-conserving constrained transport
approach based on finite-volume discretization, a linearized Riemann 
solver with parabolic reconstruction, an ability to treat multi-species
flows, a Boris-type particle pusher. In the new version of the code,
the numerical scheme and parallelization technique are advanced, 
allowing for better performance, stability, and energy conservation.
The recent improvements include: \\
(i) an account for the non-zero electron pressure; \\
(ii) an advanced particle pusher with time-centered electromagnetic 
fields (instead of previously used ``current advance method'' 
\cite{Matthews94} combined with a simple particle mover based on Taylor expansion); \\
(iii) a triangular-shaped cloud (TSC) particle model used instead of 
the nearest-grid point (NGP) model for charge deposit and 
force interpolation \cite{Birdsall,Hockney}; \\
(iii) 2d parallelization in physical space with adaptive sizes of areas, 
assigned to particular MPI processes; \\
(iv) adaptive time steps.

The structure of the paper is as follows.
Equations governing dynamics of the ions and the electromagnetic field
are discussed in Section~\ref{eqs}.
The overall numerical approach is outlined in Section~\ref{na} and briefly compared 
with the approaches of codes {\sl dHybrid} \cite{Gargate_2007} and {\sl Pegasus} \cite{Kunz2014}.
The technique and efficiency of code parallelization is illustrated 
in Section~\ref{pt}, where some estimates of computational resources needed 
to reach the scales required for typical astrophysical applications 
are given. It is shown in Section~\ref{aa} how {\sl Maximus}
can be used for modeling of particle acceleration in the Solar wind.
The simulated particle spectra agree with in-situ measurements
of the AMPTE/IRM interplanetary mission \cite{Ellison_1990}.

\section{Dymanics of the Ions and Evolution of the Electromagnetic Field} 
\label{eqs}

A standard set of equations used for hybrid modeling of ion dynamics
and evolution of the electromagnetic field (see, e.g., \cite{Matthews94}) 
can be formulated as follows: 
\begin {eqnarray}
	&&\frac{d \vec r_k}{dt}=\vec V_k \label {drdt} \\
	&&\frac{d \vec V_k}{dt}=\frac{Z_k}{m_k}\left(\vec E + \vec V_k \times \vec B \right) \label {dvdt} \\
	&&\frac {\partial \vec B}{\partial t} = - \bigtriangledown \times \vec E \label {rotE} \\
	&& \vec E = \frac{1}{\rho_c} \left(\bigtriangledown \times \vec B \right) \times \vec B 
                  - \frac{1}{\rho_c} \left( \vec j_{\rm ions} \times \vec B \right ) - \bigtriangledown P_e / \rho_c \label {elfield} \\
        && \vec j_{\rm ions} = \sum_{\rm cell} S(\vec r_k) Z_k V_k \label{j}, \hspace{3mm}
        \rho_c = \sum_{\rm cell} S(\vec r_k) Z_k 
\end {eqnarray}
Here $\vec r_k$ and $\vec V_k$, $Z_k$ and $m_k$ are positions, velocities, charges,
and masses of individual ions, $\vec E$, $\vec B$ denote the electric and magnetic field vectors, 
$\rho_c$ and $j_c$ are the ion charge and the density of the electric current integrated
over a cell of the numerical grid. The latter are derived from positions of the ions via a 
weighting function $S(\vec r_k)$. The electron pressure $P_e$ is derived from an
isothermal gyrotropic Maxwellian (this can be justified by expanding the electric 
Vlasov-Landau kinetic equation by the square root of the electron-to-ion mass ratio 
\cite{Rosin_2011}). Similar treatment of the electrons is employed in {\sl Pegasus}. 
In the early versions of {\sl dHybrid} the electron pressure is neglected, later versions
include it in the adiabatic limit \cite{caprioli_2014a}. Actually, both the adiabatic and isothermal 
approaches are somewhat artificial and seem to have little impact on the ion scales.
Some recent developments consider finite-mass electrons \cite{Amano_2014}, which is more physical
and will be implemented in {\sl Maximus} mainly for stability reasons.

The generalized Ohm's law (equation \ref{elfield}) can be represented as:
\begin{equation}
     E_j =  - \frac{1}{\rho_c} \frac {\partial P_{ij}} {\partial x_i} 
            - \frac{1}{\rho_c} \left( \vec j_{\rm ions} \times \vec B \right)_j 
\label {elfieldP}, 
\end{equation}
where $P_{ij} = (B^2/2 + P_e)\, \delta_{ij}  - B_iB_j$ is the pressure
tensor. 

These equations are further normalized by a number of natural scale units: 
the proton inertial length $l_i = c \sqrt{m_p / 4 \pi n_0 e^2}$, 
the strength of the unperturbed large-scale magnetic field $B_0$, 
the inverse proton gyrofrequency $\Omega = eB_0/m_p c$,
and the Alfven speed $V_A = B_0(4 \pi \rho_0)^{-1/2}$ 
($n_0$ and $\rho_0$ are the unperturbed plasma number and mass density respectively \footnote{
{\sl Maximus} allows treatment of multiple sorts of ions. 
However, the normalization units are always defined for the ``reference'' 
pure hydrogen plasma of the same initial mass density $\rho_0$ as the particular 
modeled composition.}).

\section{The Numerical Approach}
\label{na}

To model evolution of the ions from some initial state in a predefined spatial
area, equations (\ref{drdt}--\ref{j}) are numerically solved (advanced in time)
on a three-dimensional Cartesian grid via the standard loop of procedures,
which consists of the following steps:
\begin{enumerate}
 \item Moment Collector (calculate charges and currents from equations \ref{j})
 \item Field Solver (solve equations \ref{rotE} and \ref{elfield} 
 to follow the evolution of the electromagnetic field)
 \item Particle Mover (solve equations \ref{drdt} and \ref{dvdt} 
 to move individual ions through the grid cells)
\end{enumerate}
These steps are described in more detail below.

\subsection{Moment Collector}

At the Moment Collector step the velocities and positions of all
particles are used to calculate charge and current densities in each of
the grid cells. 

According to the standard technique of particle-in-cell modeling, 
a ``macro-particle'' approximation is used, with the macro-particles
occupying certain areas in the discrete phase space, i.e., representing
an ensemble of real (``physical'') particles, which move together. 
This approximation is inevitable due to a huge difference 
between the number of particles in real physical systems and the maximum
number of simulated particles allowed by the available computational 
resources. 
The main reason to use the finite-sized macro-particles with charge 
and mass distributed over multiple cells is suppression 
of the strong numerical noise caused by the limited number of particles
per grid cell ({\sl ppc}). The shapes of the macro-particles are defined 
by the weighting function $S(\vec r_k)$ (see equations \ref{j}). 
The impact of a weighting function on characteristic collision 
and heating time was studied by \cite{Hockney}. 
Compared to the case of point-like particles, with a second-order accurate 
triangular-shaped-cloud (TSC) weighting function we achieve an order of magnitude 
improvement of energy and momentum conservation, which could not be achieved by 
any reasonable increase of {\sl ppc}. 
On the other hand, finite-sized particles can 
suppress the instabilities with wavelengths less than the particle size. 
However, the TSC length is only twice that of a cell and 75\% of its weight 
is contained within one cell length, 
so the minimal wavelengths are still close to the grid resolution. Similar
weighting is employed in the hybrid codes {\sl dHybrid} and {\sl Pegasus}.
The TSC function is used to deposit charge and current 
densities (hereafter ``cell moments'') onto the mesh.
The cell moments are further used to update the values of the electromagnetic 
field at the Field Solver step of the algorithm.

\subsection{Field Solver}

In order to keep the divergence of magnetic field vector equal to zero,
a Godunov-type constrained transport
scheme \cite{Balsara2004} has been implemented. It employs the staggered
grid with edge-centered edge-aligned electric fields and face-centered
face-normal magnetic fields. Additional time splitting is used
to simultaneously solve the interconnected equations \ref{rotE} and
\ref{elfield}. Advancing physical time (with cell moments kept constant)
requires the values of electric and magnetic field to be leap-frogged 
several times in the following way:

\begin{enumerate}
 \item Spatial derivatives of the face-centered values of the magnetic 
 field are found. In order to keep the total variation diminishing (TVD) 
 condition, the monotonized central-difference limiter is used at this step
 (see, e.g., \cite{LeVeque97}).
 \item A piecewise-parabolic reconstruction of magnetic field inside
 each cell is made, employing the values of the normal component of 
 ${\vec B}$ and their spatial derivatives. 
 The reconstruction coefficients are calculated according to \cite{Balsara2004}.
 \item A linearized Riemann solver \cite{Balsara1998} is used to
 obtain ${\vec E}$ on the cell edges from equation \ref{elfield}. 
 \item The face-centered magnetic field is updated from equation \ref{rotE}.
\end{enumerate}

In the middle of this step the time- and cell-centered values of
$\vec E$ and $\vec B$ are obtained, to be further used at the 
Particle Mover step. For this the face-centered values of the pressure tensor $P_{ij}$
are calculated with a linearized Riemann solver near the cell faces. 
The electric field in the cell center is then found from equation \ref{elfieldP} 
via a numerical differentiation of $P_{ij}$. At the same time,
the magnetic field is found via cell-averaging of the current parabolic 
reconstruction. 
Compared to {\sl Maximus}, {\sl dHybrid} seems to lack a constrained 
transport algorithm, while in {\sl Pegasus} the cell-centered values of the 
magnetic field are simple face averages, and the evaluation of edge electric fields 
is not specified \cite{Kunz2014}.

\subsection{Particle Mover}

A Boris-type particle pusher \cite{Birdsall,Lipatov} is used to propagate
the ions. The electromagnetic field at each particle position is
calculated by interpolation of time- and cell-centered field values with
the same weighting function as during the Moment Collector step. This
ensures the second order accuracy in time and space together with the absence
of self-forces. To increase the accuracy, the following predictor-corrector 
algorithm is used:

\begin{enumerate}
\item The ``initial'' Lorentz force is calculated from current particle
positions and velocities using field interpolation.
\item The ``predicted'' particle positions and velocities at half
time-step are obtained from equations \ref{drdt} and \ref{dvdt}.
\item The ``predicted'' Lorentz force is found from ``predicted'' particle
positions and velocities.
\item The particles are moved from the ``predicted'' to final positions
using the ``predicted'' Lorentz force.
\end{enumerate}

\subsection{Implementation}

The code is written in 
C++. 
The following structure types are used: PARTICLE, SORT, CELL, FACE, EDGE, 
VERTEX. As particles are extremely memory-consuming due to their large number, 
each PARTICLE structure contains only coordinates and velocity components, but 
neither charge, nor mass. Particles of the same sort are organized in lists
inside each cell of the grid, so that they are promptly accessible. Their
charge, mass, and abundance are stored only once in a separate SORT structure.
Cell-centered field values and cell moments are stored in CELLs, while
FACEs contain values of ${\vec B}$ normal components and $P_{ij}$ components,
EDGEs contain the values of the electric field. VERTEXes are used to store some 
temporary intermediate values, which are used at the Field Solver step.

\begin{figure}
\includegraphics[width=\textwidth]{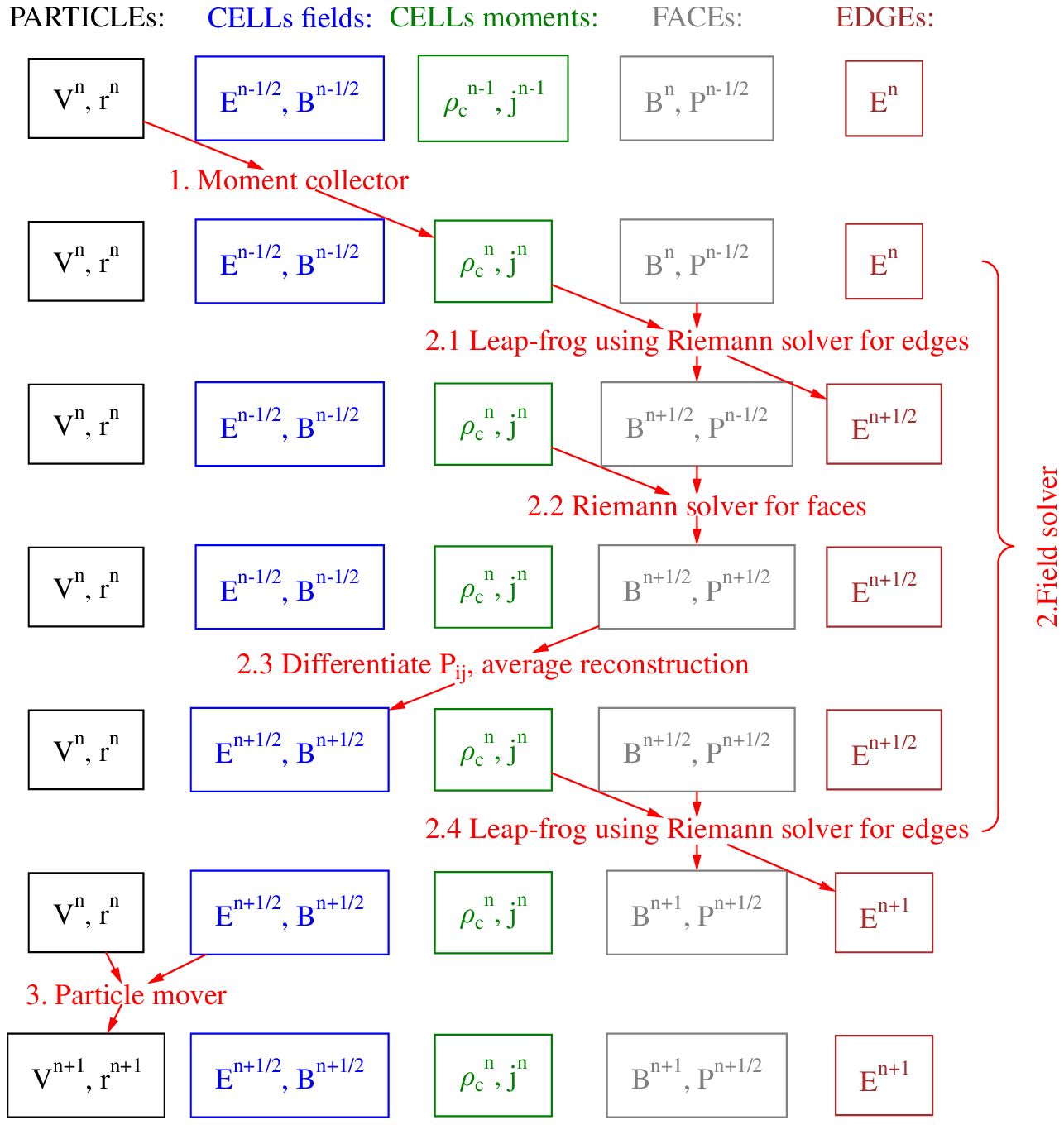}
\caption{A schematic illustration of the algorithm employed to update 
the variables from $t = t^n$ to $t = t^{n+1}$ (see Section~\ref{na} for a detailed 
description). The variables are shown as 5 color-coded categories. 
The red arrows indicate the substeps performed.
} 
\label{fig1}
\end{figure}

The numerical scheme of {\sl Maximus} is illustrated in figure~\ref{fig1}. 
All the variables are divided into 5 color-coded categories,
corresponding to the structures mentioned above. CELLs moments
and fields are separated for clarity. 

Each time step starts at $t^n$ with known particle positions and normal 
components of field vectors. First cell moments for the same moment of time
are calculated at the Moment Collector step (substep 1). Then edges of 
${\vec E}$ and faces of ${\vec B}$ are leap-frogged to half-timestep via the 
constrained transport algorithm (substep 2.1). Next, the same algorithm is 
used to find the half-timestep pressure tensor at cell faces (substep 2.2) 
and the cell-centered fields are stored (substep 2.3). Then the edges of ${\vec E}$ 
and the faces of ${\vec B}$ are leap-frogged until the end of the time step (substep2.4). 
Finally, the particles are moved to their new positions (substep 3), and 
the next step $t^{n+1}$ starts. 
Due to a similar field solver, the scheme of {\sl Pegasus} is much the same, 
though {\sl Maximus} takes the advantage of just one Particle Mover and Moment Collector 
per step, at the same time keeping the time-centering scheme.

The size of a time step can be automatically adjusted to comply with 
the numerical stability criteria \cite{Gargate_2007} and the Courant condition, 
for the particles not to cross more than one cell during one time step. 
Such an adaptive time step seems to be surprisingly absent in other
publicly known hybrid codes, though it can lead to a substantial speed-up of 
the computations.

The program sequence is realized by means of a sorted list,
which allows one to flexibly change the set of 
procedures
(for example, switch off the Moment Collector and Field Solver to
study test particle movement in external fields).

{\section{Parallelization technique and efficiency}}
\label{pt}

Usual tasks for hybrid simulations are highly resource-intensive, as the modelled
systems must be resolved on scales of ion inertial lengths and time periods less 
than their inverse gyrofrequencies. At the same time, sizes and lifetimes of typical
astrophysical objects, like supernova remnants, are many orders of magnitude larger. 
Hence hybrid simulations of such objects as a whole are not realistic.
Even though only a tiny part of the real object is modelled, its size and modeling time 
still must be substantial to study long-wave, slowly growing instabilities, 
transport of energetic particles, and their back-reaction on the background plasma. 

Another crucial parameter is the {\sl ppc} number. 
The numerical noise of the electromagnetic field values, 
which is due to the limited number of macro-particles, 
typically scales as $\delta B/B \sim ppc^{-1/2}$. 
Hence {\sl ppc} typically has to exceed 100 to guarantee 
the accuracy of simulation at least at the 10\% level.
Simulation of some finer effects may demand the {\sl ppc} 
to exceed several thousand (\cite{Florinsky_2016,Niemec_2016,kropotina_2018}).
The effects of limited {\sl ppc} on simulations of astrophysical shocks 
were investigated in \cite{kropotina_2016}. 
It should be noted that sometimes hybrid simulations of space plasmas are
performed at {\sl ppc} = 8 or even {\sl ppc} = 4 \cite{Gargate_2007,caprioli_2014a}. 
However, such simulations obviously require an artificial suppression of 
the numerical noise via low-pass filtering.
Great care should be taken in such cases, as conservation of energy and spectra
of the electromagnetic field are likely violated.

Due to their high resource-intensity, hybrid codes can be efficiently
run only on multi-core computers and clusters. The Message Passing Interface (MPI) 
parallelization technology was chosen for {\sl Maximus}: different MPI processes operate
on different spatial parts of the simulation box. 
Various tests have shown that the inclusion of the third spatial dimension is usually less
important for the modelled physical systems than maintaining substantial simulation box 
sizes in 2d. So {\sl Maximus} is usually run within 2d boxes, though all the three 
components of velocity and field vectors are considered. 
Hence, the division of the modelled physical area between the MPI processes is 
done in two dimensions. 

The directions of such division are automatically chosen along the two maximum 
simulation box sizes for each particular run. The number of processes per 
each direction is adjusted so that each area is made square-like. Notably, when
modeling a shock, the area of $r$ times higher density (i.e., higher {\sl ppc}) 
appears downstream ($r$ is the shock compression, which equals 4 for a typical 
strong shock in a monoatomic gas). Hence, the MPI regions in the shock downstream 
are made 4 times smaller to balance the load. When modeled shock is launched
at $t=0$ and then moves across the simulation box, the downstream region grows with time. 
In this case the processes domains are apriori made 4 times smaller in the region, 
which will be in downstream at the end of simulation.

\begin{figure}
\includegraphics[width=\textwidth]{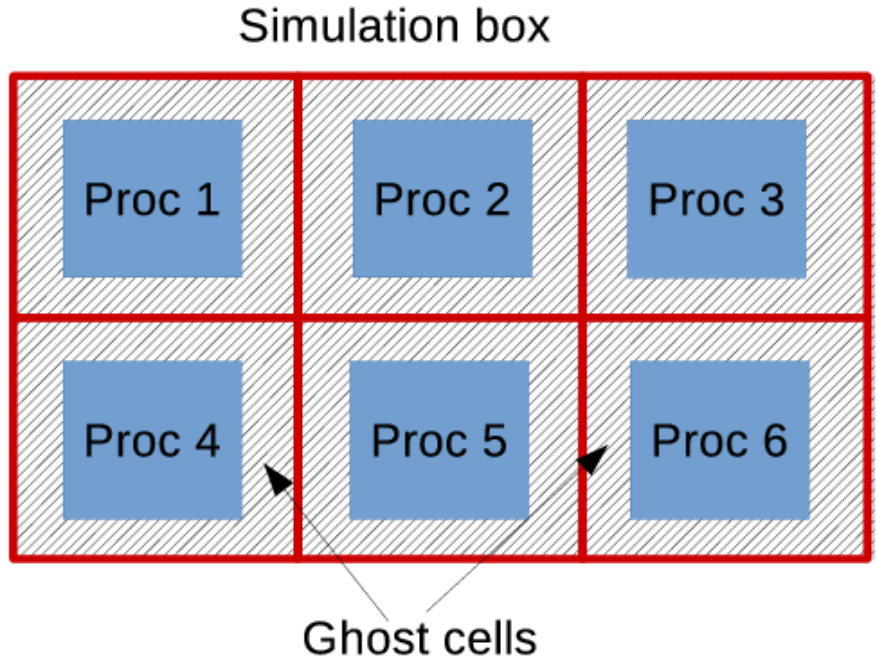}
\includegraphics[width=\textwidth]{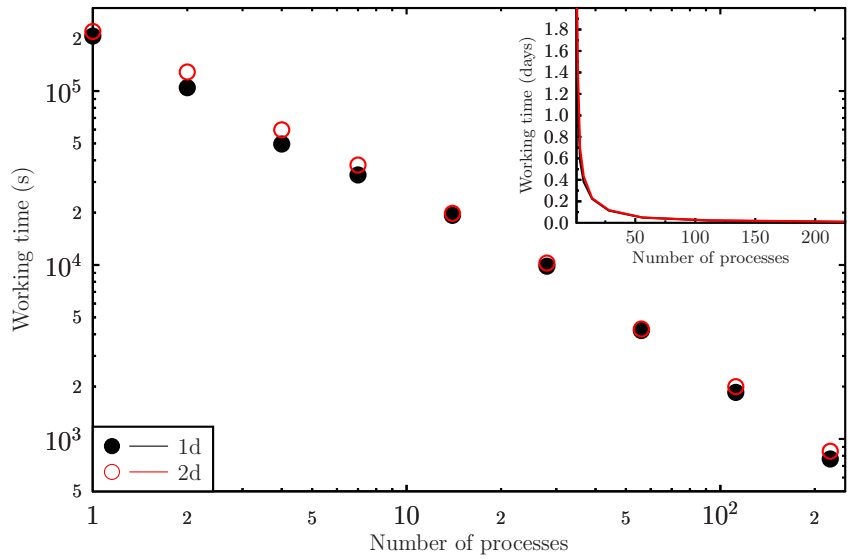}
\caption{{\bf Upper panel:} {\sl Maximus} parallelization scheme. 
The area allocated for each of the cores in the physical space
is shown in blue, while ghost cells used for interconnection 
are hatched. 
{\bf Lower panel:} Scalability of {\sl Maximus}. Run time of
the same tasks in 1d and 2d for different numbers of employed
processor cores. The top-right inside panel shows the same time, measured in days, with linear scale.
} 
\label{fig2}
\end{figure}

The numerical scheme assumes that each cell influences not only its 26 
closest 3d neighbours (for implementation of TSC particles), but also the neighbours 
of neighbours (to interpolate forces at the predicted positions of particles  
during the Particle Mover step, see Section~\ref{na}). So the spatial region 
of each process is surrounded by two layers of ghost cells, used for processes communication 
(see~Fig.~\ref{fig2}). One layer of ghost faces is also introduced for the Field Solver step. 
It should be noted that in the case of small spatial regions number of ghost cells exceeds number of
cells inside the domain, and memory and time for communication become relatively significant. 
However, this is usually not the case in large astrophysical setups. 

In order to access the parallelization efficiency, an illustrative setup 
of {\sl Maximus} was run on several sets of processes (from 1 to 28x8=224) 
on the ``Tornado'' cluster of SPbPU, which has 612 28-core nodes, 
each with 64~Gb RAM and 2 x Intel Xeon CPU E5-2697 v3 (2.60GHz).
The setup was chosen large enough to eliminate relative communication expenses.
Also the box size was taken to be divisible by the number of cores per node, so that the processes
treated equal space regions. In total, 28000 spatial cells were modelled 
with single sort {\sl ppc} = 100 and periodic boundary conditions applied 
for particles and fields.
The Moment Collector, Field Solver, and Particle Mover procedures
were executed with fixed time step $dt =$ 0.005 $\Omega^{-1}$ until
$t =$ 1000 $\Omega^{-1}$ (i.e., 200000 time steps were made). 
Two simulation box shapes were chosen: 1d (28000$\times 1 \times$1 cells)
and 2d (280 $\times 1 \times $100 cells). 

Fig.~\ref{fig2} shows the decrease
of total run time with the increase of the processes number.
One can see that the run time scales well with the number of processes, 
even if more than one node is employed. This indicates the relatively small 
contribution of internode communications to the total resource consumption, which is
expected for large box sizes. Similar estimates for {\sl dHybrid} \cite{Gargate_2007}
(with box size of 96 cells) showed substantial communication expenses due to small 
domain sizes. The difference between the 1d and 2d configurations is not substantial, 
though 1d appears slightly quicker. Overall, the characteristic run time per 
particle per step is almost constant and can be estimated as $t_{ps} \sim$ 500~ns.

Consider a typical problem of high energy astrophysics -- 
simulation of particle injection and acceleration at collisionless 
shocks formed in energetic space objects, which harbour powerful outflows
(supernova remnants, pulsar wind nebulae, active galactic nuclei, etc.). 
The number of spatial cells required to appropriately describe such an object, 
can be of order 10$^8$. If at least two sorts of ions (say, hydrogen and helium) are taken 
into account, about 100~{\sl ppc} are need, i.e., ten billion particles
are required. Being distributed among 100 nodes
(2800 cores), it averages to about 350000 particles per MPI process. Hence, one
step would take about 0.2~s, which translates to about twenty-four hours
for 500000 time steps. In fact, even such number of steps would not be enough 
to model a complex nonlinear configuration, because much smaller time steps 
may be required in that case. One may conclude, that most of the relevant 
astrophysical problems to be investigated with {\sl Maximus} demand substantial 
computational power for relatively long periods of time (from weeks to months). \\

\section{Simulation of a Collisionless Shock in the Solar Wind}
\label{aa}

\begin{figure}
\includegraphics[width=\textwidth]{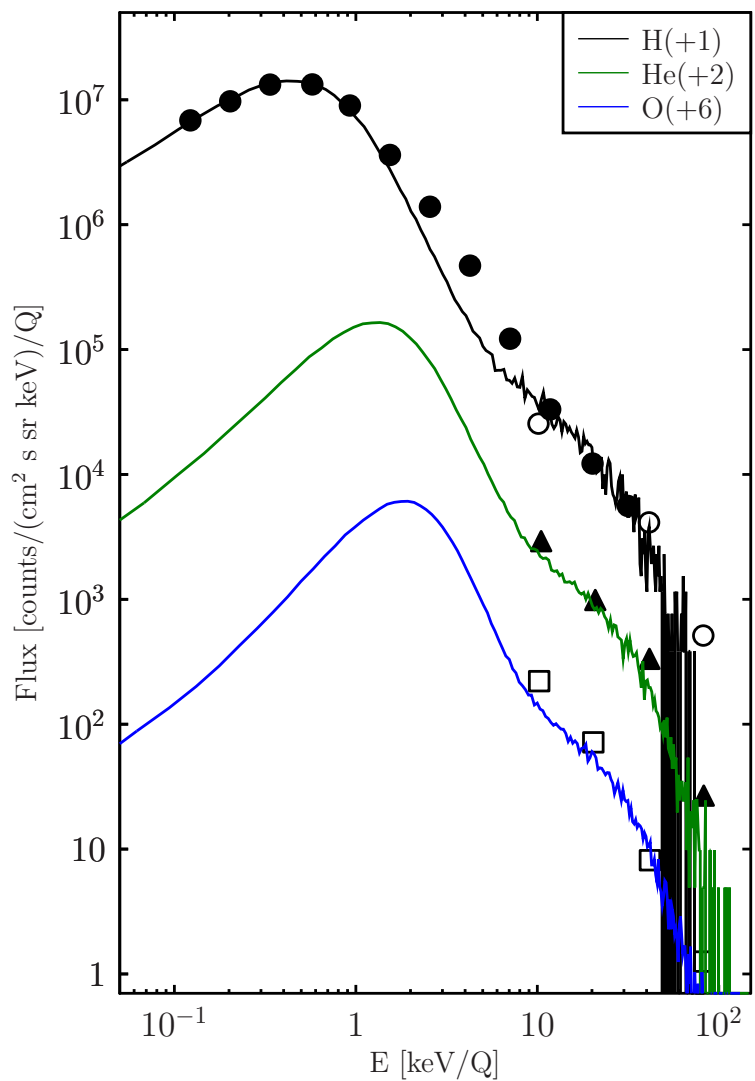}
\caption{Simulated energy distributions of ions downstream 
a shock in the Solar wind confronted with in-situ measurements
of AMPTE/IRM interplanetary mission \cite{Ellison_1990}.
}
\label{fig3}
\end{figure}

The obvious way to verify numerical models is to confront their
results to experimental data. In order to test the predictions
of {\sl Maximus}, we simulated distributions of accelerated ions
in the Solar wind and compared them with the in-situ measurements
made with the AMPTE/IRM interplanetary mission \cite{Ellison_1990}.
The simulation was run with the following parameters, taken from the direct observational data: the upstream 
Mach number (in the downstream frame) $M_A =$3.1, the magnetic field 
inclination angle $\theta =$ 10$^o$, the ratio of thermal to magnetic
pressure $\beta = $0.04. The standard Solar wind composition was considered:
90.4\% H (+1) (by mass), 8.1\% He (+2), and about 1.5\% of heavy ions,
represented by O(+6). A shock was initialized in a rectangular box sized
$10000 \times 400 l_i$ via the ``rigid piston'' method (reflection 
of a superalfvenic flow from a conducting wall). The simulated spectra
of ions in the shock downstream are shown in Fig.~\ref{fig3} together 
with the experimental data. A good agreement of the model and the observations 
can be seen; the small differences may be due to variations of shock parameters.
The acceleration Fermi process is illustrated in the supplementary video, where individual H (+1) (white circles) and He (+2) (green circles) ions
positions are placed on the protons $x-E$ phase spaces, evolving with time. Both sorts of ions are reflected
during the first shock encounter and subsequently gain energy by scattering upstream and downstream of the shock.

\section{Conclusions}
\label{cc}

The 3d second-order accurate in time and space divergence-conserving 
hybrid code {\sl Maximus} is presented. 
The code has been verified to keep energy and momentum conservation in modelled systems.
It can be efficiently run on multi-core computers and clusters due to implementation 
of the MPI parallelization technology. 
For an illustrative case of the Solar wind, 
a good agreement of the modeling with the observational data of AMPTE/IRM 
has been demonstrated. 
{\sl Maximus} can be effectively used to model nonlinear energetic processes in 
astrophysical plasmas on relevant spatial and time scales once substantial numerical 
resources are provided (hundreds of computational cores for time periods from weeks 
to months).

\section{Acknowledgments}

Some of the presented modeling was performed at the ``Tornado'' subsystem
of the supercomputing center of St.~Petersburg Polytechnic University, whose
support is readily acknowledged.

\end{document}